%% file: main.tex
\documentclass[aps, prb, reprint, twocolumn, superscriptaddress, footinbib]{revtex4-2}

\input{preamble}

\def\mytitle{Mesoscopic Heat Multiplier and Fractionalizer} 

\begin{document}

	
	\author{Florian St{\"a}bler}
		\affiliation{D\'epartement de Physique Th\'eorique, Universit\'e de Gen\`eve, CH-1211 Gen\`eve 4, Switzerland}
	\author{Eugene Sukhorukov}
		\affiliation{D\'epartement de Physique Th\'eorique, Universit\'e de Gen\`eve, CH-1211 Gen\`eve 4, Switzerland}
	
	
	\title{\mytitle}
	\date{\today}

\begin{abstract}

Local measurements of heat flux in Quantum Hall devices can deviate from the expected equilibrium heat flux due to interactions. We present a model of a simple mesoscopic device consisting of an Ohmic reservoir contacted by chiral edge states. In contrast to the well studied heat Coulomb blockade (HCB) effect, we report the opposite phenomenon, an enhancement of the heat flux carried by an edge state in the HCB regime due to an additional contribution of  the collective charge mode via the fluctuating potential of the Ohmic contact. We discuss the thermometry of these correlared states and discuss their detectability. The enhancement effect is also reflected in modified correlation functions, which influences the electrical and thermal linear response coefficients in a tunneling probe measurement. On a technical level, we introduce a Langevin formalism that elucidates the role of these extra fluctuations in electrical and thermal transport, both in uniformly heated and Joule heated devices and argue that our approach has advantages in the latter scenario compared to the standard $P(E)$ theory. We report a violation of the Wiedemann-Franz law, which is modified by the external resistance of the mesoscopic circuit. 
\end{abstract}

\maketitle

\section{Introduction}

The Ohmic contact, a small piece of metal on a highly doped semiconductor region containing a two-dimensional electron gas, is a fundamental component in electron quantum optical experiments. It provides low-resistance contact between edge states and external circuits and  control of dephasing \cite{neder_controlled_2007,roulleau_tuning_2009,roulleau_direct_2008,huynh_quantum_2012,bieri_finite-bias_2009,litvin_edge-channel_2008,litvin_decoherence_2007,helzel_counting_2015,neder_unexpected_2006} and energy equilibration \cite{altimiras_tuning_2010,altimiras_non-equilibrium_2010,le_sueur_energy_2010} through strong tunneling. Additionally, it serves as an incoherent source and detector of quasi-particles.

Despite the difficulty of providing a theoretical, microscopic description of an Ohmic contact, progress has been made by considering an effective description in terms of charge and neutral degrees of freedom \cite{slobodeniuk_equilibration_2013}. The resulting physics is rich and can be attributed to different regimes of macroscopic quantities such as the effective temperature or the charging energy of the contact. Notably, it involves dynamical  Coulomb blockade \cite{duprez_dynamical_2021,altimiras_dynamical_2014} in and out of equilibrium, heat Coulomb blockade effect \cite{sivre_electronic_2019,sivre_heat_2018}, Luttinger liquid physics \cite{anthore_circuit_2018,jezouin_tomonagaluttinger_2013,anthore_universality_2020}, charge fractionalization \cite{idrisov_quantum_2020,morel_fractionalization_2022}, and Kondo physics\cite{iftikhar_two-channel_2015,iftikhar_tunable_2018,kamata_fractionalized_2014,berg_fractional_2009,beri_exact_2017,landau_charge_2018,inoue_charge_2014}.

Our recent work focused on using an Ohmic contact as a toy model to  investigate the influence of edge reconstruction or disorder on thermal transport in quantum Hall samples \cite{stabler_transmission_2022}. We applied a combination of Langevin equations and scattering theory to a Caldeira-Leggett type system to effectively model dissipation in chiral systems. We proved the quantization of heat for these systems and derived a low energy theory for the collective degrees of freedom of the edge capturing the universal properties of the edge in the presence of dissipation.

In a subsequent step, we showed how to break the universal quantization of heat flux in non-chiral systems \cite{stabler_nonlocal_2023}, which led to the appeareance of a negative heat drag effect due to the presence of extra correlations of the collective charge degree of freedom in the edge. Our main result is that a local measurement of heat can reveal a nontrivial amount of heat current carried by edge states in a globally equilibrium system. This led to the present paper in which we analyze the electrical and thermal properties of these correlated states in a similar device.

\begin{figure}[htbp]
	\centering
	\includegraphics[width=0.7\linewidth]{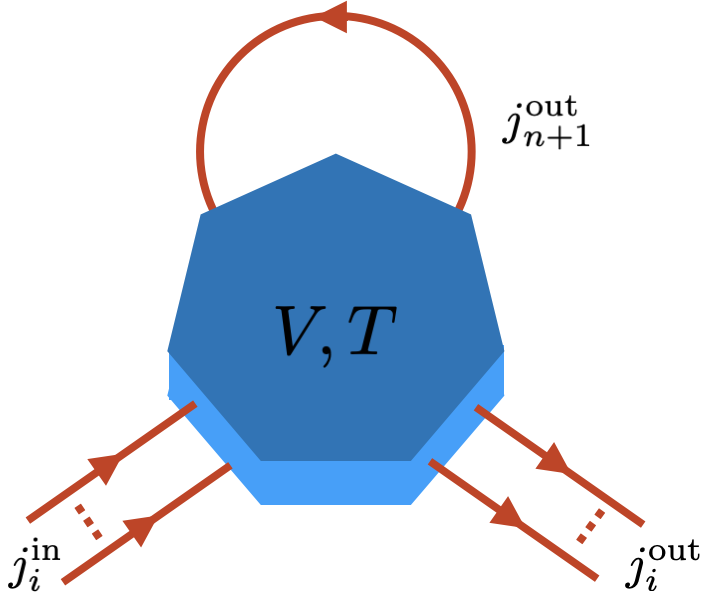}
	\caption{An Ohmic contact connected to $n$ external edge states and one looped edge state. This setup can be generalized to multiple loops. A setup like this was already experimentally realized in \cite{sivre_electronic_2019,sivre_heat_2018}. The loop can be achieved by a quantum point contact (QPC) in the pinch off regime. As we will show in \cref{sec:WF} at finite transmission the QPC can be used to probe the edge state.  We consider the Ohmic contact at a uniform temperature $T$, but experimentally the contact might be heated via Joule heating. This situation can be treated under the assumption of local thermal equilibrium by taking different temperatures in the noise powers of the boundary currents and Langevin sources, see \cref{sec:noneqL} }
	\label{fig:loopsetup}
\end{figure}

\textit{In this paper,} we present a minimal model of an Ohmic contact, and show that it can host correlated states that emerge due to strong Coulomb interactions similar to the considerations in \cite{idrisov_quantum_2020,morel_fractionalization_2022}, see \cref{fig:loopsetup}. The Ohmic contact is connected to macroscopically large electrodes via $n$ chiral edge states. One additional edge states is prepared such that it forms a loop and is fed back into the Ohmic contact. The anomalous state appears in the looped edge state. The corresponding experimental setup  closely resembles the one presented in \cite{sivre_heat_2018,sivre_electronic_2019}. We show that these correlated states are related to voltage fluctuations of the ohmic contact, which are in full agreement with P(E) theory \cite{duprez_dynamical_2021}, however the role of these extra fluctuations in electrical and thermal transport remains elusive. In the approach presented in this paper we  explicitly resolve the temperature dependence of the additional fluctuations, that had to be found with an experimental hypothesis in P(E) theory. We report a multiplication of heat, counterintuitively in the Coulomb blockade regime; A higher than quantum amount of heat carried by the interacting edge states, which poses a paradox of having a hot channel in a globally equilibrium system. The paradox is resolved by coupling this interacting state to another state  capacitively or via a quantum point contact (QPC) and explicitly showing energy conservation.  We study the linear electrical conductance and thermal tunneling conductance and report a violation of the Wiedemann-Franz law by a number that only depends on the external resistance $R = \frac{R_q}{n}$  of the ohmic contact $\mathcal{L}_n = 3\frac{2+n}{2+3n} \mathcal{L}_0$,with the resistance quantum $R_q = \frac{h}{e^2}$. Furthermore, we study the Lorenz  number in experimentally more realistic situations and numerically for arbitrary temperatures in equilibrium.

\section{Multiplication of Heat carried by an Edge State}

 Let us consider a generalization of the previously proposed model consisting of $N$ incoming and $N$ outgoing edge channels to the Ohmic contact. The bosonic Hamiltonian of this system can be written as 

\begin{equation}
    \mathcal{H} = \frac{\hbar v_F}{4 \pi} \sum_{i=1}^{2 N}\int_{-\infty}^\infty \mathop{dx} \left(\partial_x \phi_i(x,t)\right)^2 + \frac{Q^2(t)}{2C},
\end{equation} where $i=1,\dots,N$ labels the incoming, and $i=N+1,\dots,2 N$  the outgoing edge states and $Q(t) = \sum_{i'} \int_{-\infty}^0 \mathop{dx} \rho_{i'}(x,t) e^{\frac{\varepsilon x}{v_F}}$ is the integral of the charge density $\rho_i(x,t)=\frac{e}{2\pi}\partial_x \phi_i(x,t)$ inside of the interaction region $x\in (-\infty,0]$. The exponentially decaying term arises due to the finite lifetime of excitations $\varepsilon$ inside of the Ohmic contact. A similar Hamiltonian was used in the references \cite{matveev_coulomb_1995,furusaki_theory_1995,slobodeniuk_equilibration_2013}. The Heisenberg equation of motion $\partial_t \phi_i(x,t)=-\frac{i}{\hbar} \left[\phi_i(x,t),\mathcal{H}\right]$ can be recast into yet another form 

\begin{flalign}
    \frac{d}{dt}Q(t) &= \sum_{j=1}^N  \left(j^{\text{in}}_{i}(t) - j^{\text{out}}_{i}(t) \right),\label{eq:Kirchhoff}\\
    j^{\text{out}}_{i} &= \frac{1}{R_q C} Q(t) + j^{\text{c}}_{i}(t),\label{eq:Langevin}
\end{flalign} where \cref{eq:Kirchhoff} is Kirchoff's law describing the rate of change of charge on the Ohmic contact in terms of incoming and outgoing boundary currents and \cref{eq:Langevin} is a Langevin equation that contains a contribution from the collective charge mode via the fluctuating potential $ Q(t)/ C$ of the Ohmic contact and the neutral thermal current fluctuations $j^{\text{c}}_{n}(t)$.

Next, consider taking $m$ outgoing states and loop them back onto the ohmic contact, which reduces the number of unlooped external  states to $n=N-m$ incoming and outgoing states, as depicted in \cref{fig:loopsetup} \footnote{These states are referred to as external, since the series resistance of the circuit is given 
 by $R=R_q/n$.}. This can be implemented by an additional constraint on the equation of motion  \cref{eq:Kirchhoff,eq:Langevin} of the form 

\begin{equation}
    j^{\text{in}}_{k} = e^{i  \frac{\omega L}{v_F}}  j^{\text{out}}_{k} \quad n < k  \leq N  \label{eq:LoopCond},
\end{equation} 

Our goal is to compute the heat flux carried by the looped edge states, which is expected to be anomalous, since additionally to the free fermionic neutral excitation carrying a heat flux quantum, we expect the presence of some charge fluctuations adding to this heat flux.

To compute the heat flux carried by any edge state outgoing of the ohmic contact, we solve \cref{eq:Kirchhoff,eq:Langevin,eq:LoopCond} and express  the outgoing currents in terms of incoming  boundary currents and the Langevin sources. For simplicity let us assume, the system is in thermal equilibrium, i.e. the noise power of the incoming boundary current  and the noise power of the Langevin sources are equal to an equilibrium noise power at a known temperature $S_{\text{in}}=S_{\text{c}}= S_{\text{eq}}$.

\begin{equation}\label{eq:Seq}
    S_{\text{eq}} = \frac{1}{R_q}\frac{\hbar \omega}{1-e^{-\beta \hbar \omega}},
\end{equation} were $\beta = \frac{1}{k_B T}$ is the inverse temperature. This constraint will be relaxed in \cref{sec:noneqL}.  Furthermore let us assume we look at the simple case of one loop $m=1$ and $n$ external channels and short loops $L=0$, which corresponds to the experimental situation in \cite{sivre_heat_2018}. For more general solutions see \cref{app:GeneralCase}.

\begin{figure}[htbp]
    \centering
    \includegraphics[width=1\linewidth]{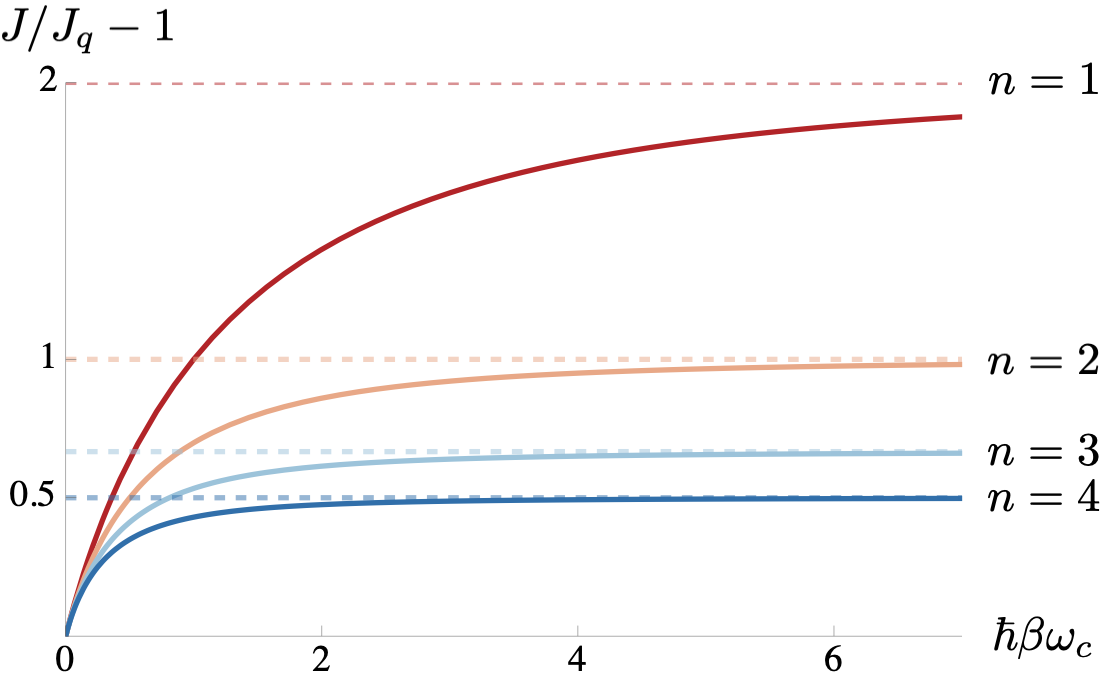}
    \caption{The excess heat flux in the looped edge state normalized to a quantum $J/J_q -1$ as a function of the dimensionless inverse temperature $\hbar \beta \omega_c \sim E_c / T$ for $ n \in \{1,2,3,4\}$ (solid lines) external edge channels. Note that increasing the number of edge channels decreases the magnitude of the excess heat flux as well as increasing the convergence towards the strongly interacting low temperature limit (dashed lines).  }
    \label{fig:HFgenRCandn}
\end{figure}

To obtain the heat carried by an outgoing edge state we integrate the noise power over all frequencies  \footnote{The expression for the heat flux used in this paper is motivated from a continuity equation argument, see \cite{stabler_transmission_2022}. Note that the heat flux can sometimes be defined as the integral of the broadening of the electron distribution function $J=\int \mathop{d\varepsilon} \varepsilon (f(\epsilon)-\theta(-\epsilon))$. Since the dispersion relation of the edge states is linear outside of the ohmic contact these to definitions are the same, see \cite{levkivskyi_energy_2012}.}.

\begin{equation} \label{eq:Heat}
	J = R_q\int_{-\infty}^\infty \frac{d\omega}{4\pi} \left\{S_i(\omega) - \left.S_i(\omega)\right|_{T \rightarrow 0}\right\},
\end{equation} where the noise power is defined as

\begin{equation}
S_{i}(\omega) = 2\pi \delta(\omega+\omega') \left\langle \delta j^{\text{out}}_{i}(\omega)\delta j^{\text{out}}_{i}(\omega') \right\rangle,
\end{equation} and can be found by a Fourier transformation of  \cref{eq:Kirchhoff,eq:Langevin,eq:LoopCond}. The noise power is given by

\begin{gather}
S_{i} (\omega)=  S_{\text{eq}}(\omega), \quad   1 < i \leq n ,\\
S_{n+1}(\omega)= \left(1+\frac{2 n}{n^2+ (\omega/\omega_c) ^2}\right) S_{\text{eq}}(\omega),\label{eq:Prefactor}
\end{gather} where the time constant $\omega_c^{-1} = R_q C = \frac{\pi \hbar}{E_c}$ is related to the charging energy $E_c =\frac{e^2}{2 C}$. The expression for the noise power of the external channels is equilibrium as expected from the unitarity of the scattering matrix, however the noise power of the loop acquires an additional contribution due to the charge fluctuations of the Ohmic contact. According to \cref{eq:Heat} the heat flux carried by the looped edge state is given by
\begin{equation}\label{eq:Jfree}
    J = \frac{\pi}{12 \hbar } k_B^2 T^2 =   J_q, 
\end{equation}%
for high temperatures compared to the charging energy $ \beta E_c  \ll 1$ and 
\begin{equation}
    J =  \left(1+\frac{2}{n} \right) J_q, \label{eq:Jloop}
\end{equation}%
for low temperatures $   \beta E_c \gg 1$. Let us stress again, that the anomalous heat flux can be understood as a contribution from the fermionic neutral mode and charge fluctuations created by the Ohmic contact. Solving \cref{eq:Heat} numerically allows to explore the intermediate values of $J$ as a function of the channel number $n$ and the dimensionless parameter $\hbar \beta \omega_c \sim E_c/ T$. The graph in \cref{fig:HFgenRCandn} shows the dependence of the heat flux for different values of the dimensionless inverse temperature and number of channels.

\section{Edge State Thermometry - Capacitive Coupling}

Imagine the following setup, where the looped edge state is capacitively coupled to another edge state. If the device is in global thermal equilibrium, i.e. the temperature of the noise power of the incoming currents and sources are all equal to the base temperature, the heat flux carried by the looped channel is given by \cref{eq:Jloop} and could also be understood in terms of an effective temperature of the channel $T_{\text{eff}} \sim \sqrt{1+\frac{2}{n} } T$ that is higher than the base temperature. In a normal resonator a temperature difference would lead to a net energy flux \cite{sukhorukov_resonant_2007}, however here this enhancement is due to charge fluctuations of the Ohmic reservoir. Nevertheless, we need to address this paradox, since it is impossible to extract energy from the system in global equilibrium according to the second law of thermodynamics.

\begin{figure}[htbp]
    \centering
    \includegraphics[width=.6\linewidth]{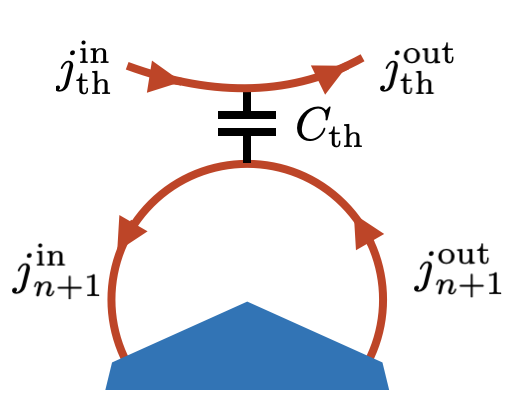}
    \caption{Capacitive resonator coupling the looped edge state to a free edge state. The thermometer can be operated in equilibrium or with a temperature difference between the edge states. }
    \label{fig:setupC}
\end{figure}

Adding the capacitively coupled channel as a thermometer can be implemented by adding the following equations to \cref{eq:Kirchhoff,eq:Langevin,eq:LoopCond}

\begin{equation}
    \begin{pmatrix}
    j^{\text{out}}_{\text{th}}\\
    j^{\text{in}}_{n+1}
    \end{pmatrix} = \begin{pmatrix}
    r_{\text{th}} & t_{\text{th}}\\
    t_{\text{th}} & r_{\text{th}}
    \end{pmatrix} \begin{pmatrix}
    j^{\text{in}}_{\text{th}}\\
   j^{\text{out}}_{n+1}\end{pmatrix},
\end{equation} where the transmission and reflection amplitudes of the thermometer are given by

\begin{gather}
    t_{\text{th}}=\frac{1-2 e^{i \omega t_W}+e^{2i \omega t_W}}{2-2 e^{i \omega t_W}-i  \omega/\omega_\text{th} },\\
    r_{\text{th}}=\frac{1-e^{i\omega t_W}+i e^{i\omega t_W}  \omega / \omega_\text{th} }{2-2 e^{i\omega t_W}+i   \omega/\omega_\text{th} },
\end{gather} where $t_W= W /v_F$ is the time of flight through the resonator and $\omega_\text{th}^{-1} = R_q C_{\text{th}}$ is the RC-time of the capacitive interaction strength between the thermometer and the interacting mode, see \cref{app:Thermometer}.

In a next step we compute the heat flux in the outgoing part of the thermometer to check if heat is carried over from the hot channel to the cold channel, using the same protocol described earlier we evaluate \cref{eq:Heat}. Due to the unitarity of the scattering matrix for the device and thermometer we find the expected result 

\begin{equation}
    S_{\text{th}}(\omega)= 2\pi  \left\langle \delta j^{\text{out}}_{\text{th}}(\omega)\delta j^{\text{out}}_{\text{th}}(-\omega) \right\rangle = S_{\text{eq}},
\end{equation} which yields a heat flux quantum according to the equivalent expression to \cref{eq:Heat}. This resolves the paradox, since no heat can be extracted if the system is in equilibrium and the additional correlations adding to the heat flux exactly cancel in the computation of the outgoing heat flux of the thermometer.

It is also possible to study non-equilibrium situations for small or large $W$ or for small frequencies. Let us assume a temperature of the thermometer that is different from the equilibrium temperature, but let us keep the assumption of local thermal equilibrium. We can define the noise power of incoming current fluctuations to be of equilibrium form but with a different temperature $S^{\text{in}}_{\text{th}}(T_\text{th}) = S_{\text{eq}}(T_\text{th})$ with $T_\text{th}>T_0$. If we assume the distance $W$ on which the thermometer interacts with the system is short $\frac{k_B T t_W}{\hbar} \ll 1$ we find

\begin{multline}
    S_{\text{th}}(\omega) = S^\text{in}_{\text{th}}(\omega)  \\ \hspace{-.2cm} + \!\! \left(\!1\!+\!\frac{2n}{n^2 \! + (\omega / \omega_c)^2} \right) \! \omega^2  \omega_\text{th}^2   t_W^4 \! \left( S_{\text{eq}}(\omega) \!-\!  S^\text{in}_{\text{th}}(\omega)\right)\!,
\end{multline} which appears as $T^4$ correction to the heat flux after taking the heat flux integral \cref{eq:Heat}. One can see that the presence of the correlated state can change the (non-universal) prefactor of the correction in contrast to the non-interacting case, but not the overall scaling. In the limit of $W\rightarrow \infty$ the reflection and transmission amplitudes resemble the ones of the Ohmic contact, which is a limit that is interesting to study, but will be considered elsewhere.

\section{Breakdown of Wiedemann-Franz law}
\label{sec:WF}

\subsection{Lorenz number of tunneling current}
\begin{figure}[htbp]
    \centering
    \includegraphics[width=.6\linewidth]{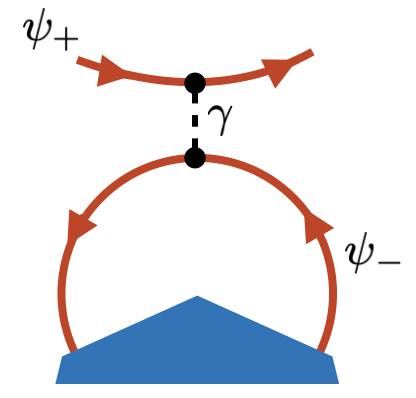}
    \caption{We describe the QPC using the tunneling Hamiltonian formalism. An electron is created in the plus edge and annihilated in the minus edge, or vice versa with a perturbatively small interaction constant $\gamma$. The tunneling is switched on adiabatically, which allows us to treat it separately from the Coulomb interactions in the Ohmic contact. We compute the electrical and thermal conductance to leading order in the tunneling constant.}
    \label{fig:setupQPC}
\end{figure}

Let us assume a coupling of a looped  edge state to a free one via a quantum point contact (QPC) to leading order of tunneling, see \cref{fig:setupQPC}. In this system we also expect no energy transport in equilibrium, but additionally, we report the violation of Wiedemann-Franz law by a universal number that depends only on the number of external channels, but not on microscopic details of the Ohmic contact, the edge states or the QPC itself. To do this we employ the tunneling Hamiltonian approach, which allows the computation of both, electrical conductance and thermal conductance.

Let us consider the Hamiltonian of the QPC $\mathcal{H} = \sum_{\sigma=\pm} \mathcal{H}_\sigma + \mathcal{H}_T$, which consists of two counter propagating channels connected by a tunneling term with 

\begin{gather}
    \mathcal{H}_\sigma = - i \sigma \hbar v_F   \int_{-\infty}^\infty \mathop{dx} \psi^\dagger_\sigma(x,t) \partial_x \psi_\sigma(x,t),\\
    \mathcal{H}_T = \gamma \psi^\dagger_+(x_0,t) \psi_-(x_0,t) +  \text{h.c.}.
\end{gather}

This formalism treats  tunneling separately from the interactions at the Ohmic contact, which allows us to impose fermionic boundary conditions for the free edge state $\psi_+$ and non-trivial interacting boundary conditions due to interactions at the Ohmic contact. For detailed calculations see \cref{app:tunneling}. The average current and average heat flux are defined by the rate of change of charge  and energy  of the free system respectively. This gives for the average tunneling current

\begin{multline}\label{eq:Itun}
      I  = 
   \frac{e\gamma^2}{\hbar^2}  \int_{-\infty}^\infty \mathop{d t}  \left( \left\langle \psi_+(t) \psi_+^\dagger(0)\right\rangle \left\langle \psi_-^\dagger(t) \psi_-(0)\right\rangle \right. \\- \left.\left\langle \psi_+^\dagger(0) \psi_+(t)\right\rangle \left\langle \psi_-(0) \psi_-^\dagger(t)\right\rangle\right),
\end{multline}
and  for the tunneling heat flux
\begin{multline} \label{eq:Jtun}
      J  = \frac{ \gamma^2 }{i \hbar} \int_{-\infty}^0 \!\! \mathop{dt} \left( \left\langle   \psi^\dagger_-(0) \psi_-(t) \right\rangle \left\langle \dot{\psi}_+(0)  \psi^\dagger_+(t)   \right\rangle \right.  \\ \left.
      -  \left\langle   \psi_-(t)  \psi^\dagger_-(0)\right\rangle \left\langle  \psi^\dagger_+(t)   \dot{\psi}_+(0) \right\rangle \right),
\end{multline} where $\dot{\psi}(t') = \lim\limits_{t \rightarrow t'}\partial_t \psi(t)$.

We evaluate these expressions using the non-equilibrium bosonization technique \cite{sukhorukov_scattering_2016}, which allows us to express the vertex operators $\psi_\sigma(t) = \frac{1}{\sqrt{2 \pi a}}   \exp\left( i \phi_\sigma(t) \right) $ in terms of bosonic fields $\phi_\sigma(t) $, which can be found from the currents \cref{eq:Langevin}. Every two-point function can be expressed due to the Gaussian nature of the theory as

\begin{gather} \label{eq:cf1}
    \left\langle \psi_\sigma(t) \psi_\sigma^\dagger(t')\right\rangle = e^{-i\mu_\sigma(t-t')} \mathcal{C}_\sigma(t-t'),\\\label{eq:cf2}
    \left\langle \psi^\dagger_\sigma(t) \psi_\sigma(t')\right\rangle = e^{i\mu_\sigma(t-t')} \mathcal{C}_\sigma(t-t'),
\end{gather} where the first exponential function is the zero mode contribution and the phase correlation function $\ln \mathcal{C}_\sigma(t) = \left\langle(\phi_\sigma(0)-\phi_{\sigma}(t))\phi_{\sigma}(0)\right\rangle$ is related to the noise power via

\begin{equation}  \label{eq:cfi}
 \ln \mathcal{C}_\sigma(t) = - \frac{2\pi}{e^2}\int \frac{{}\mathop{d\omega}}{\omega^2} S_{\sigma}(\omega) (1-e^{-i\omega t}),
\end{equation} where $S_+(\omega) = S_{\text{eq}}(\omega)$ and $S_-(\omega) = S_{n+1}(\omega)$. The free fermionic phase correlation function is given by the standard expression

\begin{equation}\label{eq:cffree}
    \mathcal{C}_+(t) =-\frac{i}{2\hbar \beta  v_F} \frac{1}{\sinh\left(\frac{\pi}{ \hbar \beta }\left(t - i \eta\right)\right)},
\end{equation} where the shift of the pole was introduced to find the correct Fourier transformation in terms of the occupation number. For the interacting phase correlation function we find it by relying on a separation of energy scales. The correlation function is altered by the interaction parameter $\omega_c$. For temperatures much larger than the charging energy $\hbar \omega_c \ll k_B T$, we know that our correlation function must have free fermionic character, e.g. due to \cref{eq:Jfree}. For temperatures much smaller than the charging energy, which is in turn much smaller than the UV cutoff, we can find an asymptotic solution in the long time limit $\omega_c t \gg 1$ of the form 

\begin{multline}\label{eq:cfint}
     \mathcal{C}_-(t)=     \frac{\left( \frac{    \pi e^{-\gamma_{\text{EM}}} }{  \omega_c n }  \right)^{\frac{2}{n}} }{2  v_F} \\\times \left(\frac{  1}{ i\hbar \beta } \frac{1}{\sinh\left(\frac{\pi }{ \hbar \beta}\left(t-t'-i\eta\right)\right)} \right)^{1+\frac{2}{n}}\hspace{-.7cm},
\end{multline} see \cref{app:CF}.

This allows us to compute  \cref{eq:Jtun,eq:Itun}, which yield very similar expressions, except for the time derivative in the heat flux. Let us start from the average current. We can compute a $\frac{dI}{dV}$, for small bias, which gives. Details of the calculations can be found in \cref{app:avgIavgJ};  we will  give the results assuming a separation of energy scales given by $k_B T ,\hbar t^{-1}  \ll \hbar \omega_c  \ll \hbar v_F a^{-1} $.
\subsubsection{Linear conductance}

The linear conductance is given by

\begin{equation}
    G_n = \frac{e^2   }{2 \pi  \hbar 
     } \frac{ \gamma ^2  }{ \hbar^2 
   v_F^2  }\left( \frac{    \pi e^{-\gamma_{\text{EM}}} }{\hbar \beta \omega_c n }  \right)^{\frac{2}{n}} \frac{ \sqrt{\pi} \Gamma \left(1+\frac{1}{n}\right)}{2 \Gamma \left(\frac{3}{2}+\frac{1}{n}\right)}, 
\end{equation} which in general depends on the number of channels $n$ and reproduces the free fermionic result

\begin{equation}
    G_\infty= \frac{e^2   }{2 \pi \hbar 
   } \frac{ \gamma ^2  }{ \hbar^2 
   v_F^2  }.
\end{equation}

\subsubsection{Heat flux}

It is easy to see that the heat flux is zero for equal temperatures of the two terminals. For a small temperature difference $\delta T$ the heat flux is given by

\begin{multline} \label{eq:Jn}
      J_n  =   \frac{4^{\frac{1}{n}} \left((n-1) \Gamma \left(2+\frac{1}{n}\right)^2-\Gamma \left(1+\frac{1}{n}\right) \Gamma \left(3+\frac{1}{n}\right)\right)}{n \Gamma
   \left(4+\frac{2}{n}\right)} \\
   \times \frac{ \pi  k_B  \delta T }{  \beta  \hbar } \left( \frac{    \pi e^{-\gamma_{\text{EM}}} }{\hbar \beta \omega_c n }  \right)^{\frac{2}{n}}\frac{ \gamma^2 }{ \hbar^2 v_F^2 },
\end{multline} Especially the non interacting case gives

\begin{equation}
    J_\infty =  \frac{ \pi  k_B  \delta T }{ 6 \beta  \hbar }   \frac{ \gamma^2 }{ \hbar^2 v_F^2 }.
\end{equation}

The Lorenz number for a specific number of channels $n$ is given by

\begin{equation}
    \mathcal{L}_n = \frac{J_n k_B \beta  }{\delta T G_n}  =  \frac{2+n }{2+3 n} \frac{ \pi^{2}  k_B^2 }{e^2  } = 3\frac{2+n }{2+3 n} \mathcal{L}_\infty.
\end{equation}

Note that the same result may be obtained by considering transport integrals \cite{karki_coulomb_2020,kiselev_generalized_2023}.

\subsection{Lorenz number for selective temperature bias} \label{sec:noneqL}

In real experiments e.g. in the experiments where dynamical Coulomb blockade is studied with a temperature bias \cite{duprez_dynamical_2021,sivre_electronic_2019}, the Ohmic contact is heated by Joule heating. This may lead to a non-uniform temperature in which the base temperature of the device and the temperature of the sources of the reservoir can be different. In that case the noise power in the loop is not given by \cref{eq:Prefactor}, but instead by

\begin{equation} \label{eq:SldT}
S_l(\omega) = S_c(\omega ) +   \frac{n \left( S_c(\omega ) + S_{\text{in}}(\omega )\right) }{n^2+ (\omega/\omega_c) ^2},
\end{equation} where the noise power is given by \cref{eq:Seq} with the respective temperatures  $T_c$ of the Langevin sources and $T_{\text{in}}$ for the current fluctuations incident to the device. 

In \cite{duprez_dynamical_2021}, it was stated that after the Joule heating the temperature of the node is different from the temperature of the electromagnetic environment. The best fit to the data was given by the mean temperature of the $(T_{\text{in}}+T_c)/2$ with small deviations. \Cref{eq:SldT} shows that this conjecture is valid to leading order in small temperature differences between node and environment and corrections appear in second order of $T_{\text{in}}-T_c$. This is a strong advantage of the non-equilibrium bosonization approach \cite{sukhorukov_scattering_2016}, which allows to have a more intuitive physical picture, by describing the voltage fluctuations in terms of boundary currents. This allows to derive the aforementioned temperature of the voltage fluctuations rigorously.

The nontrivial temperature dependence of the noise power translates directly to the phase correlation function and thus the Lorenz number. Imagine only heating the Ohmic contact and keeping the other channels at base temperature, i.e. $T_{\text{in}}=T_+ = T$. This would alter the correlation function \cref{eq:cfint} to

\begin{multline}
     \mathcal{C}_-(t)=     \frac{\left( \frac{    \pi e^{-\gamma_{\text{EM}}} }{  \omega_c n }  \right)^{\frac{2}{n}} }{2  v_F} \\\times \left(\frac{  k_B T}{ i\hbar  } \frac{1}{\sinh\left(\frac{\pi k_B T }{ \hbar }\left(t-t'-i\eta\right)\right)} \right)^{\frac{1}{n}} \\\times \left(\frac{  k_B T_c}{ i\hbar  } \frac{1}{\sinh\left(\frac{\pi k_B T_c }{ \hbar }\left(t-t'-i\eta\right)\right)} \right)^{1+\frac{1}{n}} ,
\end{multline}

where we assume $T_c=T+\delta T$ and repeat the calculations leading to the heat flux \cref{eq:Jn}. This leads to a modified Lorentz number given by

\begin{multline}
    \frac{\mathcal{L}_n^{\text{sel.}}}{\mathcal{L}_\infty} = \frac{3 n \Gamma \left(\frac{3}{2}+\frac{1}{n}\right)}{ \sqrt{\pi }(2 n+1)  \Gamma
   \left(1+\frac{1}{n}\right)} \\\times \left(\, _2F_1\left(1,-2-\frac{1}{n};2+\frac{1}{n};-1\right)-1\right) ,
\end{multline} where $_2F_1$ is the hyper geometric function. The maximum Lorenz number is reduced from $\mathcal{L}_1 = \frac{9}{5}$ to $\mathcal{L}_1^{\text{sel.}} = \frac{6}{5}$.

\subsection{Temperature dependence of the Lorenz number}

If the separation of energy scales $k_B T ,\hbar t^{-1} \ll \hbar \omega_c \ll  \hbar v_F a^{-1} $ is not fulfilled, it is possible to study the exact solution for linear conductance and thermal conductance numerically \footnote{We assume that the tunneling strength $\gamma$ is the smallest energy scale. Otherwise other pertubations might become relevant. See the discussion in \cite{kane_thermal_1996}.}. In the following we plot the Lorenz number for a device with  $T_{\text{in}}=T_c$ computed from the numerical solutions of \cref{eq:Itun,eq:Heat} using the exact correlation function, see \cite{duprez_dynamical_2021}, which after shifting the poles similar to before gives the following Lorenz number

\begin{equation} \label{eq:lnum}
    \frac{\mathcal{L}_n}{\mathcal{L}_\infty} = \frac{\displaystyle\int_0^\infty \!\!\! \mathop{dx} \ \frac{3}{2} (3-\cosh (2 x)) \text{sech}^4(x)e^{\Phi(x)}}{\displaystyle\int_0^\infty \! \!\! \mathop{dx} \ \text{sech}^2(x)e^{\Phi(x)}},
\end{equation} with the hyperbolic secant function $\text{sech}(x)=(\cosh(x))^{-1}$.

\begin{widetext}
\begin{multline}
    \Phi(x) = -\frac{2 x}{n}+\frac{\pi \left( \cos \left(\frac{n \hbar \beta  \omega_c}{2}  \right)-e^{-\frac{n x \hbar \beta  \omega_c}{\pi }}\right)}{n \sin
   \left(\frac{n  \hbar \beta  \omega_c }{2} \right)}-\frac{2 \log \left(1+e^{-2 x}\right)}{n} \\- \frac{1}{n} \sum_{\sigma = \pm 1} H_{\frac{\sigma n  \hbar \beta  \omega_c   }{2
   \pi }}+e^{- i \frac{\sigma n  \hbar \beta  \omega_c   }{2
    } +\frac{ \sigma n x \hbar \beta  \omega_c }{\pi }  } B\left(-e^{-2
   x};1+\frac{\sigma  n\hbar \beta  \omega_c}{2 \pi },0\right),
\end{multline}
\end{widetext} where $H_\alpha$ is a harmonic number and $B(x;a,b)$ is the incomplete Beta function.

\begin{figure}[htbp]
    \centering
    \includegraphics[width=.8\linewidth]{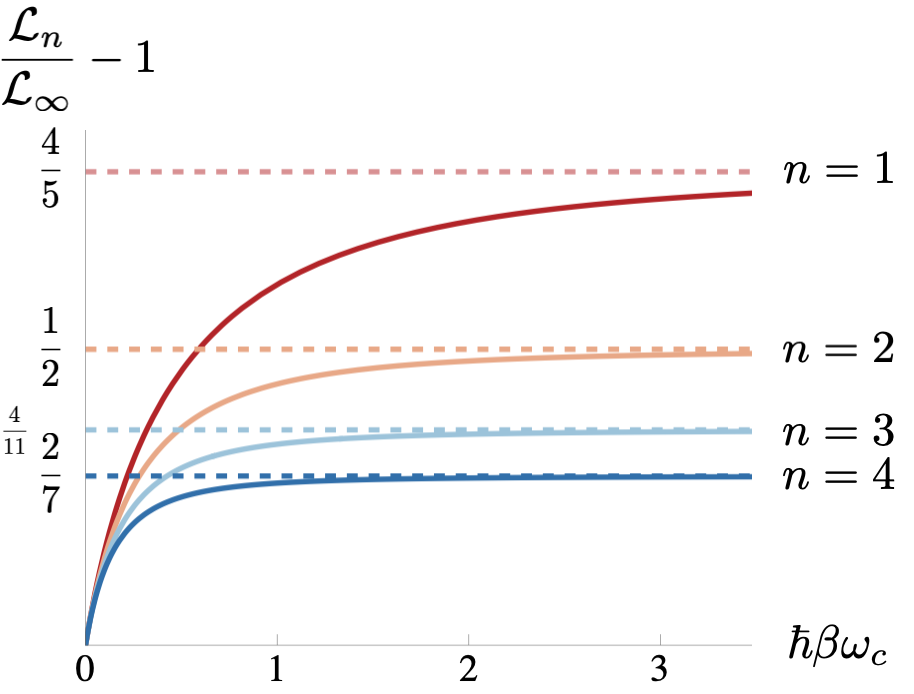}
    \caption{Numerical values of the Lorenz number for $n\in \{1,2,3,4\}$ (solid lines). The dashed lines indicate the theoretical predictions for the low-temperature, strongly interacting regime}
    \label{fig:lnumerics}
\end{figure}

 The ratio of the Lorenz number minus one is plotted in \cref{fig:lnumerics}.
 
Note that a large number of channel suppresses the magnitude of the Lorenz number, i.e. the effect of the voltage fluctuations, but also leads to a faster convergence towards the strongly interacting prediction indicated by the dashed lines upon lowering the temperature. The Lorenz number for arbitrary temperatures can be studied in a more cumbersome, but similar way by deriving the equivalent expression of \cref{eq:lnum} using \cref{eq:SldT}.

\subsection{Weak backscattering limit}

\begin{figure}[htbp]
    \centering
    \includegraphics[width=.5\linewidth]{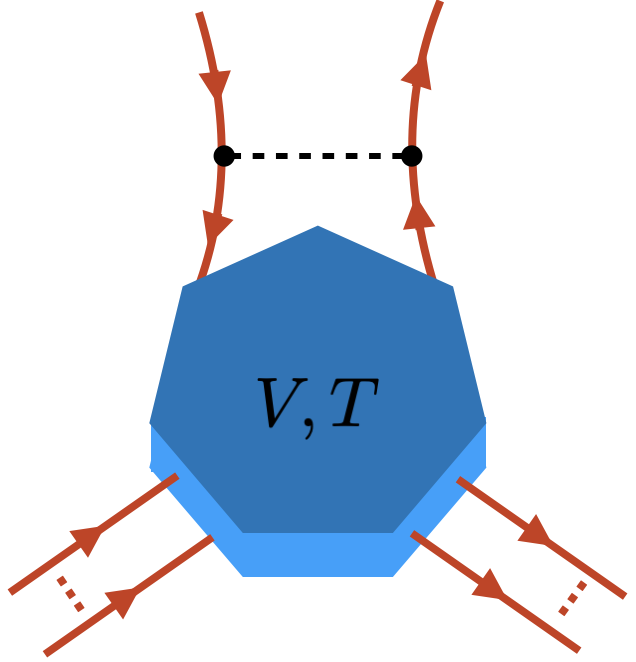}
    \caption{QPC in the weak backscattering configuration. The Lorenz number is expected to be dual to the weak tunneling case.}
    \label{fig:dual}
\end{figure}

If the QPC is operated in the opposite regime of weak tunneling, see \cref{fig:dual}, we expect a Lorenz number that is given by the mapping $n \rightarrow - (n-1)$, see \cref{appsec:dual}. This is in full agreement with the duality in the Luttinger parameter discussed in \cite{duprez_dynamical_2021}. The Lorenz number in the weak back scattering regime is given by

\begin{equation}
    \mathcal{L}_n = \frac{3 (1-n)}{2-3 (1+n)} \mathcal{L}_\infty.
\end{equation}

\subsection{Additional comments}

The Lorenz number presented in this paper was calculated in an experimentally established setup, where the external resistance can only be changed by changing the series resistance of the Ohmic contact in discrete steps $R = R_q/n$, however one can also imagine a circuit in series with a macroscopic resistor that can be freely tuned. The computations in this paper do not rely on $n$ being integer, which allows us to immediately generalize our result for an arbitrary series resistance $R$ for the Lorenz number

\begin{equation}
     \mathcal{L}_n  = 3\frac{2+R_q/R }{2+3 R_q/R } \mathcal{L}_\infty.
\end{equation} As was shown in \cite{duprez_dynamical_2021,anthore_circuit_2018,jezouin_tomonagaluttinger_2013,safi_one-channel_2004} there is a direct  mapping between  an electronic channel in series with a linear
resistance and  a Tomonaga-Luttinger model (TLL) for an
infinitely long 1D system of spinless electrons interacting with a single impurity. The interaction parameter of the TLL is given by 
\begin{equation}
    K = \left(1+ \frac{R}{R_q}\right)^{-1},
\end{equation} which extends our results directly to the TLL case mentioned above. 

A violation of the Wiedemann-Franz law was also reported in interacting Luttinger liquids by \cite{kane_thermal_1996}, but the reported dependence of the prefactor on the interaction parameter is different.

\section{Conclusion}

In this paper, we provided new insights into the previously overlooked heat multiplication effect within Ohmic contacts. The enhancement of locally measured heat flux appears in the Coulomb blockade regime, which at first seems counter intuitive, but is due to the presence of an additional  contribution
from the collective charge mode via the fluctuating
potential $Q(t)/C$ of the Ohmic contact.  We used a capacitive resonator and a quantum point contact to perform thermometry on this seemingly hotter edge state and ruled out any paradoxes involving the violation of energy conservation. 

One consequence of this is the violation of the tunneling  Wiedemann-Franz law leading to a new set of  universal Lorenz numbers, the ratio of linear thermal response to voltage response, that depend only on the series resistance $R$ of the circuit. We at first assume a separation of energy scales and study this Lorenz number in the Coulomb blockade regime where the charging energy is much larger than the energy scale set by temperature. We used a combination of Langevin equation and scattering theory to study a uniformly heated and, experimentally more relevant, selectively heated Ohmic contact, leading to a larger than expected Lorenz ratio for both cases. Our approach allows for a more intuitive picture of the role of fluctuations compared to P(E) theory, especially in the context of non-local heat transport discussed in \cite{stabler_nonlocal_2023}.  Namely it allows us to derive the experimental hypothesis of the environmental temperature made in \cite{duprez_dynamical_2021} rigorously.  

We conducted numerical studies to investigate the temperature dependence of the Lorenz number beyond Coulomb blockade regime and discuss how our model can be mapped to continous resitances, a TLL model and how it can be applied for fractional fillings, where this approach could be used to probe the edge structure of complicated filling factors. During the preparation of the manuscript we became aware of \cite{kiselev_generalized_2023} reporting on a similar violation of the Wiedemann-Franz law.

In summary, our findings laid the foundation for further studies on out-of-equilibrium situations involving Ohmic contacts. As an outlook, our results open the way to study this new universality in existing HCB systems. It can be generalized and adds to the understanding of thermo-electric transport in fractional quantum hall systems and Kondo circuits.

The authors acknowledges the financial support from
the Swiss National Science Foundation.

\bibliographystyle{apsrev4-2}
\bibliography{Paper.bib}

\appendix

\section{Dependence on external resistance and number of loops} \label{app:GeneralCase}

The heat flux transported inside of the loop depends on many parameters, namely, the external resistance, i.e. the number of channels attached to the Ohmic contact that do not form a loop, the length of the loop, and the number of loops. The Langevin equations for $n$ external channels and $m$ loops of length $L$ are given by

\begin{flalign}
    - i \omega Q(\omega) &= \sum_{k=1}^{n+m}
    \left(j^{\text{in}}_{k}(\omega) - j^{\text{out}}_{k}(\omega) \right), \\
    j^{\text{out}}_{k}(\omega)  &= \frac{1}{R_q C} Q(\omega) + j^{\text{c}}_{k}(\omega),\\
     j^{\text{in}}_{l}(\omega)  &= e^{i\omega t_{\text{fl}}} j^{\text{out}}_{l}(\omega), \quad l>n
\end{flalign}

\subsection{Noise power for multiple loops of length $L$ with uniform heating}
Let us assume equal temperatures for the noise powers of boundary currents and Langevin sources we find the noise power of one of the looped edge states to be $S_l(\omega) = f(\omega) S_{\text{eq}}(\omega)$, with
\begin{equation} \label{eq:GeneralCase}
  \hspace{-.15cm} f(\omega) \! =\!  1-\frac{1}{ m+\frac{2 m n+n^2+  (\omega/\omega_c) ^2}{2(m-(m+n) \cos ( \omega t_\text{fl} )+\omega/\omega_c  \sin (\omega t_\text{fl}))}},
\end{equation} with $t_{\text{fl}}=L/v_F$. The heat flux carried by the looped edge state can be found by performing the integral in \cref{eq:Heat}. If we take $L\rightarrow \infty$ a separate averaging of $f(\omega)$ over the oscillations shows that $f(\omega)\rightarrow 1$. From an experimental point of view the assumption of neglecting potential retardation effect due to a finite length of the loop is justified \cite{sivre_heat_2018}.

\subsection{ Noise power for short loops with selective heating}

In a non-equilibrium situation, where  $S_{\text{in}}\neq S_{\text{c}}$ and under the assumptions of short loops, we find

\begin{equation}
S_l(\omega) = S_{cL}(\omega ) +   \frac{n \left( S_c(\omega ) + S_{\text{in}}(\omega )\right) }{n^2+ (\omega/\omega_c) ^2},
\end{equation} which reproduces  \cref{eq:Prefactor}.

\section{Scattering matrix of the capacitive thermometer} \label{app:Thermometer}

The equation of motion for the capacitive thermometer are derived from a Hamiltonian of two counterpropagating edge states that interact within a region of size $W$ capacitively with coupling strength $C_{\text{th}}$.

\begin{equation}
    \partial_t \phi_\sigma(x,t)+ v_F \partial_x \phi_\sigma(x,t) = -\frac{e}{\hbar C_{\text{th}}} Q_{\text{th}} \theta(x) \theta(W-x), 
\end{equation} where $Q_{\text{th}}= \sum_{\sigma'} \int_0^W \mathop{dx}\ \rho_{\sigma'}(x)$ is the total charge inside of the interaction region and $\sigma=\pm$ encodes the chirality of the edge states.

\begin{figure}[htbp]
	\centering
	\includegraphics[width=0.5\linewidth]{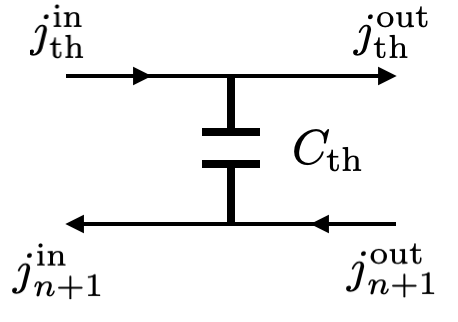}
	\caption{The labelling of currents of the thermometer.}
	\label{fig:Ccoupling}
\end{figure}

The solutions of the equations of motion for $0<x<W$ in terms of bosonic currents is given by 

\begin{equation}
    j_\sigma(x,\omega) = \Tilde{j}_\sigma(\omega) e^{\frac{i \omega  \sigma  x}{v}}+\frac{
   1-e^{\frac{i \omega  \sigma  x}{v}}}{i \omega /\omega_{\text{th} }\sigma } \sum_{\sigma'}\!\!\int_0^W \!\!dx\ \partial_x j_{\sigma'}(\omega,x),
\end{equation} to which we can assign the boundary currents

\begin{gather}
    j^{\text{in}}_{\text{th}}(\omega) \overset{!}{=} j_+(0,\omega),\\
    j^{\text{out}}_{\text{th}}(\omega) \overset{!}{=} j_+(W,\omega),\\
    j^{\text{in}}_{\text{n+1}}(\omega) \overset{!}{=} - j_-(0,\omega),\\
    j^{\text{out}}_{\text{th}}(\omega) \overset{!}{=} - j_-(W,\omega),\\
\end{gather} which allows us to solve for the S-matrix of the system which is given by

\begin{equation}
    \begin{pmatrix}
    j^{\text{out}}_{\text{th}}\\
    j^{\text{in}}_{n+1}
    \end{pmatrix} = \begin{pmatrix}
    r_{\text{th}} & t_{\text{th}}\\
    t_{\text{th}} & r_{\text{th}}
    \end{pmatrix} \begin{pmatrix}
    j^{\text{in}}_{\text{th}}\\
   j^{\text{out}}_{n+1}\end{pmatrix},
\end{equation} where the transmission and reflection amplitudes of the thermometer are given by

\begin{gather}
    t_{\text{th}}=\frac{1-2 e^{i \omega t_W}+e^{2i \omega t_W}}{2-2 e^{i \omega t_W}-i  \omega/\omega_\text{th} },\\
    r_{\text{th}}=\frac{1-e^{i\omega t_W}+i e^{i\omega t_W}  \omega / \omega_\text{th} }{2-2 e^{i\omega t_W}+i   \omega/\omega_\text{th} },
\end{gather} where $t_W= W /v_F$ is the time of flight inside of the interaction region and $\omega_{\text{th}}^{-1}=R_Q C_{\text{th}} $ is the RC-time of the capacitive interaction strength between the thermometer and the interacting mode.

\section{Tunneling Hamiltonian approach} \label{app:tunneling}

The Hamiltonian of the system is given by 

\begin{equation}
    \mathcal{H} = \mathcal{H}_+ + \mathcal{H}_- + \mathcal{H}_T,
\end{equation}
where
\begin{equation}
    \mathcal{H}_\pm = \mp i \hbar v_F  \int_{-\infty}^\infty \mathop{dx} \psi^\dagger_\pm(x,t) \partial_x \psi_\pm(x,t)
\end{equation}
and

\begin{equation}
    \mathcal{H}_T = \gamma \psi^\dagger_+(x_0,t) \psi_-(x_0,t) +  \text{h.c.},
\end{equation} where we use  $\{\psi_+(x,t),\psi^\dagger_{+}(y,t)\}= \delta(x-y)$.

\subsection{Average Heat Flux}

The heat flux operator is given by 

\begin{multline}
    \hat{J} = \frac{-i}{\hbar} \left[\mathcal{H}_+,\mathcal{H}_T\right] = - \gamma v_F \int_{-\infty}^\infty \mathop{dx} \\ \times\left[\psi^\dagger_+(x,t) \partial_x\psi_+(x,t) , \psi^\dagger_+(x_0,t) \psi_-(x_0,t) +  \text{h.c.}\right],
\end{multline} where we use the equation of motion and find

\begin{multline}
    \left[\psi^\dagger_+(x,t) \partial_t \psi_+(x,t)  ,\psi^\dagger_+(x_0,t) \psi_-(x_0,t) \right]=  \\= - \partial_t \psi^\dagger_+(x_0,t) \psi_-(x_0,t) \delta(x-x_0),  
\end{multline}

and

\begin{multline}
    \left[\psi^\dagger_+(x,t) \partial_t \psi_+(x,t) , \psi^\dagger_-(x_0,t) \psi_+(x_0,t)\right] \\=  -  \psi^\dagger_-(x_0,t) \partial_t \psi_+(x_0,t)   \delta(x-x_0).
\end{multline}

Which yields

\begin{equation}
    \hat{J}  =  \gamma  \left(  \partial_t \psi^\dagger_+(x_0,t) \psi_-(x_0,t)  +   \text{h.c.}\right)
\end{equation}

The average is given by

\begin{multline}
   J= \left\langle \hat{J} \right\rangle = - \frac{i}{\hbar} \int_{-\infty}^0 \mathop{dt} \left\langle \left[J(0), \mathcal{H}(t)\right] \right\rangle \\=  \frac{- i \gamma^2 }{\hbar} \int_{-\infty}^0 \mathop{dt} \\\times\left\langle \left[   \partial_t \psi^\dagger_+(0) \psi_-(0) +  \text{h.c.} , \psi^\dagger_+(t) \psi_-(t) +  \text{h.c.}\right] \right\rangle \\
    =\frac{- i \gamma^2 }{\hbar} \int_{-\infty}^0 \mathop{dt} \left\langle \left[   \partial_t \psi^\dagger_+(0) \psi_-(0) ,  \psi^\dagger_-(t) \psi_+(t)\right] \right\rangle \\
    +  \left\langle \left[  \psi^\dagger_-(0) \partial_t \psi_+(0) , \psi^\dagger_+(t) \psi_-(t) \right] \right\rangle\\
    =\frac{- i \gamma^2 }{\hbar} \int_{-\infty}^\infty \mathop{dt} \left\langle \left[   \partial_t \psi^\dagger_+(0) \psi_-(0) ,  \psi^\dagger_-(t) \psi_+(t)\right] \right\rangle \\
   = \frac{- i \gamma^2 }{\hbar} \int_{-\infty}^\infty \mathop{dt} \left\langle    \partial_t \psi^\dagger_+(0)  \psi_+(t) \right\rangle\left\langle   \psi_-(0)   \psi^\dagger_-(t) \right\rangle \\
    - \left\langle      \psi_+(t)  \partial_t \psi^\dagger_+(0)\right\rangle\left\langle  \psi^\dagger_-(t)  \psi_-(0)   \right\rangle
\end{multline}

\subsection{Average tunneling current}

\begin{multline}
    \hat{I} = \frac{-ie}{\hbar} \int_{-\infty}^\infty \mathop{d x}\left[\psi_+^\dagger(x,t) \psi_+(x,t),\mathcal{H}_T\right] \\=  \frac{-i \gamma}{\hbar} \int_{-\infty}^\infty \mathop{d x}\left[\psi_+^\dagger(x,t) \psi_+(x,t),\psi^\dagger_+(x_0,t) \psi_-(x_0,t)+  \text{h.c.}\right]  \\
    = \frac{-ie \gamma}{\hbar} \int_{-\infty}^\infty \mathop{d x}\left[\psi_+^\dagger(x,t) \psi_+(x,t),\psi^\dagger_+(x_0,t) \psi_-(x_0,t)\right] \\+ \left[\psi_+^\dagger(x,t) \psi_+(x,t),  \psi^\dagger_-(x_0,t) \psi_+(x_0,t)\right] 
\end{multline}

\begin{multline}
    \left[\psi_+^\dagger(x,t) \psi_+(x,t),\psi^\dagger_+(x_0,t) \psi_-(x_0,t)\right] \\= \psi_+^\dagger(x,t)\psi_-(x_0,t) \delta(x-x_0)
\end{multline}

\begin{multline}
     \left[\psi_+^\dagger(x,t) \psi_+(x,t),  \psi^\dagger_-(x_0,t) \psi_+(x_0,t)\right] \\= -\psi_+(x,t) \psi^\dagger_-(x_0,t) \delta(x-x_0)
\end{multline}

\begin{equation}
    \hat{I} = \frac{-ie \gamma}{\hbar} \left(\psi_+^\dagger(x_0,t)\psi_-(x_0,t) - \psi^\dagger_-(x_0,t) \psi_+(x_0,t)  \right)
\end{equation}

\begin{multline}
 I=  \left\langle \hat{I} \right\rangle = \frac{-i}{\hbar} \int_{-\infty}^0 \mathop{d t} \left\langle\left[I_T(0),\mathcal{H}_T\right] \right\rangle \\=-\frac{e\gamma^2}{\hbar^2}  \int_{-\infty}^0 \mathop{d t}   \left\langle \left[\psi_+^\dagger(0)\psi_-(0) , \psi^\dagger_-(t) \psi_+(t)\right]\right\rangle \\- \left\langle \left[ \psi^\dagger_-(0) \psi_+(0),\psi^\dagger_+(t) \psi_-(t) \right]\right\rangle  \\ = 
   -\frac{e\gamma^2}{\hbar^2}  \int_{-\infty}^\infty \mathop{d t} \left\langle \psi_+^\dagger(0) \psi_+(t)\right\rangle \left\langle \psi_-(0) \psi_-^\dagger(t)\right\rangle \\- \left\langle \psi_+(t) \psi_+^\dagger(0)\right\rangle \left\langle \psi_-^\dagger(t) \psi_-(0)\right\rangle ,
\end{multline} 

\subsection{Weak Backscattering limit}
\label{appsec:dual}
In the weak back scattering limit the the correlation functions of the $+$ and $-$ arm cannot be averaged separately. We thus start from the equation for current

\begin{multline}
 I=  \frac{e\gamma^2}{\hbar^2}  \int_{-\infty}^\infty \mathop{d t}   \left\langle \left[\psi^\dagger_-(t) \psi_+(t),\psi_+^\dagger(0)\psi_-(0) \right]\right\rangle 
\end{multline} 

Where we resum the exponentials due to the Gaussian nature of the theory. One of the terms in the tunneling current is given by

\begin{equation}
   K(t) = \left\langle  e^{-i\phi_-(t)}  e^{i\phi_+(t)}  e^{-i\phi_+(0)}  e^{i\phi_-(0)}\right\rangle 
\end{equation}

\begin{multline}
   -\ln K(t) =\int  \frac{\mathop{d\omega}}{\omega} \left(1-\frac{2 (n+1)}{(n+1)^2 + (\omega/\omega_c)^2}\right)\\\times\frac{1-e^{-i\omega t}}{1-e^{-\beta \hbar \omega}},
\end{multline} which implies that the current, but also heat flux and Lorenz number can be obtained from the weak backscattering case by the mapping 

\begin{equation}
    n \rightarrow -(n+1),
\end{equation} which is a result of the duality of the weak back scattering and tunneling case.

\section{(Non-) Interacting Correlation function}
\label{app:CF}

\subsection{Free fermionic correlation funciton}

\cref{eq:cfi} can be manipulated, by rewriting the Bose-function as a geometric series, respecting the sign of $\omega$, which gives

\begin{multline}
  - \ln \mathcal{C}_+(t) = \frac{2\pi}{e^2} \int  \frac{\mathop{d\omega}}{\omega} \frac{1-e^{-i\omega t}}{1-e^{-\beta \hbar \omega}} \\
  = \int_0^\infty  \frac{\mathop{d\omega}}{\omega} (1-e^{-i\omega t}) e^{- \frac{a \omega}{v_F}} \\
  + \sum_{n=1}^\infty \int_0^\infty  \frac{\mathop{d\omega}}{\omega} (2-2\cos(\omega t)) e^{-\beta \hbar \omega n}\\= \log\left( \frac{i v_F t + a}{a} \right) + \sum_{n=1}^\infty \log\left(1+ \frac{t^2}{n^2 \beta_\pm^2 \hbar^2}\right) \\
  = \log\left(\left( \frac{i v_F t + a}{a} \right)\prod_{n=1}^\infty\left(1+ \frac{t^2}{n^2 \beta^2 \hbar^2}\right)\right)\\= \log\left(\left( \frac{i v_F t + a}{a} \right) \frac{\hbar \beta}{\pi t} \sinh\left(\frac{\pi t}{\beta \hbar}\right)\right),
\end{multline} where we introduced the real space UV cutoff given by $a^{-1}$, which gives the correlation function

\begin{equation}
    \left\langle \psi^\dagger_+(t) \psi_+(t')\right\rangle =  -\frac{i}{2 \pi v_F} \frac{\pi}{\hbar \beta} \frac{e^{i\mu_+(t-t')}}{\sinh\left(\frac{\pi}{\beta \hbar} \left(t-t'-i\eta\right)\right)},
\end{equation} where the shift of the pole is chosen such that we obtain the correct Fermi distribution function if the expression is Fourier transformed and especially in the zero temperature limit $\beta \rightarrow \infty$ we find

\begin{equation}
    \left\langle \psi^\dagger_+(t) \psi_+(t')\right\rangle = -\frac{i}{2 \pi v_F}  \frac{e^{i\mu_+(t-t')}}{t-t'-i\eta}.
\end{equation} The correlation function of the conjugated therm can be found according to  \cref{eq:cf1,eq:cf2}.

\subsection{Interacting case}

We would like to find the correlation function respecting the energy following separation of energy scales $k_B T ,\hbar t^{-1} \ll \hbar \omega_c  \ll \hbar v_F a^{-1} $,
 where the time $t$ is supposed to be far from the UV regime, i.e. we consider a long time limit. This gives

\begin{multline}
  - \ln \mathcal{C}_-(t) =\int  \frac{\mathop{d\omega}}{\omega} \left(1+\frac{2 n}{n^2 + (\omega/\omega_c)^2}\right)\frac{1-e^{-i\omega t}}{1-e^{-\beta \hbar \omega}} \\= - \ln \mathcal{C}_+(t) + \int_{\frac{1}{i t}}^{\frac{v_F}{a}}  \frac{\mathop{d\omega}}{\omega} \frac{2 n}{n^2 +(\omega/\omega_c)^2}\\ 
  + \frac{2 }{n} \lim_{t\rightarrow \infty} \int_{0}^{\frac{v_F}{a}}  \frac{\mathop{d\omega}}{\omega} \left(1-e^{-i\omega t}\right)  - \log\left(i \frac{v_F}{a} t \right)\\
  + \frac{2}{n} \sum_{n=1}^\infty \int_0^\infty  \frac{\mathop{d\omega}}{\omega} (2-2\cos(\omega t)) e^{-\beta \hbar \omega n}\\= - \ln \mathcal{C}_+(t) \\+ \frac{2}{n} \left(\log\left( i n \omega_c t \right) +\gamma_{\text{EM}}+\log\left(\frac{\hbar \beta}{\pi t} \sinh\left(\frac{\pi t}{\beta \hbar}\right)\right)\right),
\end{multline} where $\gamma_{\text{EM}} \approx 0.577\dots$ is the Euler-Mascheroni constant. First note that the integral can be split into two parts: the free fermionic part and one part that contains the interaction. From now on we only consider the correction. We split the integral into two parts as before. For the temperature independent part, we cut the integral at $\omega \rightarrow \frac{1}{i t}$, since the exponential function is fast oscillating for large t. Since the integral is in principle convergent for small frequencies, we have to take into account the possibility to find a constant term that does not vanish in the long time limit. To do this we also cut the integral at high frequencies only, subtract the logarithmic divergence, which is already accounted for in the long-time limit and take $t\rightarrow \infty$ in what remains. This gives an additional constant factor.  The upper limit is given by the UV cutoff. For the temperature dependent part of the integral, we set $ \omega_c \rightarrow \infty$, since the charging energy is much larger than the energy scale set by temperature and hence
 
\begin{multline}
    \left\langle \psi^\dagger_{-}(t) \psi_{-}(t')\right\rangle =  \underbrace{-\frac{i}{2 \pi v_F} \frac{\pi}{\hbar \beta} \frac{e^{i\mu_-(t-t')}}{\sinh\left(\frac{\pi}{\beta \hbar} \left(t-t'-i\eta\right)\right)}}_{\text{free fermionic }} \\
    \times
    \left(\frac{  \pi}{i n \hbar \beta \omega_c e^{\gamma_{\text{EM}}} } \frac{1}{\sinh\left(\frac{\pi }{\beta \hbar}\left(t-t'-i\eta\right)\right)} \right)^{\frac{2}{n}},
\end{multline} and for $\beta \rightarrow \infty$

\begin{multline}
    \left\langle \psi^\dagger_{-}(t) \psi_{-}(t')\right\rangle = -\frac{i}{2 \pi v_F}  \frac{e^{i\mu_-(t-t')}}{t-t'-i\eta} \\\times \left(\frac{ 1 }{i n  \omega_c e^{\gamma_{\text{EM}}}} \frac{1}{t-t'-i\eta} \right)^{\frac{2}{n}}.
\end{multline} Note that the Euler constant appears only due to how the integral has been regularized. In this case we choose to cut-off the integral. Regularization with an exponential decay does not produce this constant as is shown for the free fermionic correlation function, but in any case it remains unphysical and should contribute to the normalization of the correlation function.

\section{Average Current and average Heat Flux} \label{app:avgIavgJ}

\subsection{Average Current}

The average current is given by \cref{eq:Itun}, where the correlation functions are given by \cref{eq:cffree,eq:cfint}, respectively. This yields the following integral
\begin{multline}
       I  = 
   \frac{e\gamma^2}{\hbar^2}  \int_{-\infty}^\infty \mathop{d t}  \left( \left\langle \psi_+(t) \psi_+^\dagger(0)\right\rangle \left\langle \psi_-^\dagger(t) \psi_-(0)\right\rangle \right. \\
   - \left.\left\langle \psi_+^\dagger(0) \psi_+(t)\right\rangle \left\langle \psi_-(0) \psi_-^\dagger(t)\right\rangle\right).
\end{multline}

At zero temperature, we note that the operator $\psi^\dagger$
applied to the ground state creates an electron-like excitation above the
Fermi level (with the positive energy), while the operator $\psi$ creates a hole-like one below Fermi level (with the negative energy). One consequence of this is that all singularities in the first term are
shifted to the upper half plane of the complex variable t, whereas they are shifted to the lower half plane in the second term. This means only one term contributes depending of the sign of the bias, i.e. depending on if we close the contour in the upper and lower half plane. At finite temperatures we also have occupied states above the Fermi level, due to thermal activation processes. We thus have to take into account both terms simultaneously. 

\begin{multline}
    I = -\left( \frac{    \pi e^{-\gamma_{\text{EM}}} }{\hbar \beta \omega_c n }  \right)^{\frac{2}{n}} \frac{e  \gamma ^2  }{4 \pi \hbar^3 \beta
   v_F^2  } \int_{-\infty}^\infty \mathop{dt} e^{i \frac{e V  \beta   }{\pi} t} \\\times \sum_{\sigma= \pm1}\frac{\sigma}{  \left(-i \sigma \sinh(t+ i \sigma \frac{\pi \eta}{\hbar \beta} )\right)^{2+\frac{2}{n}}},
\end{multline} where we  shift the contour $t \rightarrow t + i \sigma \frac{\pi}{2}$ and use that $\sinh(t+i \sigma \frac{\pi}{2}) =i \sigma \cosh(t) $ and set $\eta \rightarrow 0$. This gives 

\begin{multline}
    I = -\left( \frac{    \pi e^{-\gamma_{\text{EM}}} }{\hbar \beta \omega_c n }  \right)^{\frac{2}{n}} \frac{e  \gamma ^2  }{4 \pi \hbar^3 \beta
   v_F^2  } \int_{-\infty}^\infty \mathop{dt} e^{i \frac{e V   \beta   }{\pi} t} \\\times \sum_{\sigma= \pm1}\frac{\sigma e^{- \frac{e V   \beta   }{2} }}{ \cosh^{2+\frac{2}{n}}(t)},
\end{multline}

where we can perform the sum over $\sigma$, which gives

\begin{multline}
    I = \left( \frac{    \pi e^{-\gamma_{\text{EM}}} }{\hbar \beta \omega_c n }  \right)^{\frac{2}{n}} \frac{e  \gamma ^2  }{2  \pi \hbar^3 \beta
   v_F^2  } \sinh\left(\frac{e V   \beta   }{2} \right)\\\times
   \int_{-\infty}^\infty \mathop{dt} e^{i \frac{e V  \hbar \beta   }{\pi} t} \frac{1}{  \cosh^{2+\frac{2}{n}}(t)},
\end{multline} where the last integral can be evaluate by taking $t\rightarrow \log(z)$ and express the result in terms of Gamma functions. We find

\begin{multline}
    I = \frac{e  \gamma ^2  }{2 \pi  \hbar^3 \beta
   v_F^2  }\left( \frac{    \pi e^{-\gamma_{\text{EM}}} }{\hbar \beta \omega_c n }  \right)^{\frac{2}{n}}  \sinh\left(\frac{e V   \beta   }{2} \right) \\\times \frac{2^{\frac{2+n}{n}} \Gamma \left(1+\frac{1}{n}-\frac{i e V \beta  \hbar }{2 \pi }\right) \Gamma \left(1+\frac{1}{n}+\frac{i e V \beta  \hbar }{2 \pi }\right)}{\Gamma
   \left(2+\frac{2}{n}\right)}, 
\end{multline} 

Which gives the linear conductance

\begin{equation}
    G_n = \frac{e^2   }{2 \pi  \hbar 
     } \frac{ \gamma ^2  }{ \hbar^2 
   v_F^2  } \left( \frac{    \pi e^{-\gamma_{\text{EM}}} }{\hbar \beta \omega_c n }  \right)^{\frac{2}{n}} \frac{ \sqrt{\pi} \Gamma \left(1+\frac{1}{n}\right)}{2 \Gamma \left(\frac{3}{2}+\frac{1}{n}\right)}, 
\end{equation} which reproduces the free fermionic result

\begin{equation}
    G_\infty= \frac{e^2   }{2 \pi \hbar 
    } \frac{ \gamma ^2  }{ \hbar^2 
   v_F^2  }.
\end{equation}

\subsection{Average Heat Flux}
The average heat flux is given by a similar expression

\begin{multline}
      J  = \frac{- i \gamma^2 }{\hbar} \int_{-\infty}^\infty \mathop{dt} \left\langle   \dot{\psi}^\dagger_+(0)  \psi_+(t) \right\rangle\left\langle   \psi_-(0)   \psi^\dagger_-(t) \right\rangle \\ - \left\langle      \psi_+(t)   \dot{\psi}^\dagger_+(0)\right\rangle\left\langle  \psi^\dagger_-(t)  \psi_-(0)   \right\rangle,
\end{multline} in which the correlation functions depend on different temperatures and $\dot{\psi}(t) = \lim_{t'\rightarrow t} \partial_{t'} \psi(t')$. For equal temperatures the expression vanishes as expected from the second law of thermodynamics. We assume the temperature difference is small and expand in small $\delta T $, which gives the following heat flux.

\begin{multline}
      J  =- \frac{i }{16 n} \frac{ k_B  \delta T }{  \beta  \hbar } \left( \frac{    \pi e^{-\gamma_{\text{EM}}} }{\hbar \beta \omega_c n }  \right)^{\frac{2}{n}} \frac{ \gamma^2 }{ \hbar^2 v_F^2 } \int_{-\infty}^\infty \mathop{dt}   \sum_{\sigma=\pm 1} \left(\sigma i \right)^{\frac{2}{n}} \\\times  \frac{ 2 t \left(1-n+\cosh
   \left(2 t+\frac{2  \pi i \sigma  \eta }{\beta  \hbar }\right)\right)-(2-n) \sinh \left(2 t+\frac{2 \pi i \sigma \eta }{\beta  \hbar }\right) }{\sigma\sinh^4\left(t+\frac{\pi i \sigma  \eta }{\beta  \hbar }\right) \sinh^{\frac{2}{n}}\left(t+\frac{\pi i \sigma \eta }{\beta  \hbar }\right)} 
\end{multline} where we again shift the poles up and down similar to before and take $\eta \rightarrow 0 $, which gives

\begin{multline}
      J_n  = \frac{1}{8 n} \frac{ \pi  k_B  \delta T }{  \beta  \hbar } \left( \frac{    \pi e^{-\gamma_{\text{EM}}} }{\hbar \beta \omega_c n }  \right)^{\frac{2}{n}} \frac{ \gamma^2 }{ \hbar^2 v_F^2 } \\\times
      \int_{-\infty}^\infty \mathop{dt}  \frac{ 1-n-\cosh (2 t)}{\cosh^{4+\frac{2}{n}}(t)} \\=  \frac{ \pi  k_B  \delta T }{  \beta  \hbar } \left( \frac{    \pi e^{-\gamma_{\text{EM}}} }{\hbar \beta \omega_c n }  \right)^{\frac{2}{n}} \frac{ \gamma^2 }{ \hbar^2 v_F^2 } \\\times 
      \frac{4^{\frac{1}{n}} \left((n-1) \Gamma \left(2+\frac{1}{n}\right)^2-\Gamma \left(1+\frac{1}{n}\right) \Gamma \left(3+\frac{1}{n}\right)\right)}{n \Gamma
   \left(4+\frac{2}{n}\right)},
\end{multline} which can be evaluated as before by the transformation $t\rightarrow \log(z)$. Especially the non interacting case gives

\begin{equation}
    J_\infty =  \frac{ \pi  k_B  \delta T }{ 6 \beta  \hbar }   \frac{ \gamma^2 }{ \hbar^2 v_F^2 }.
\end{equation}

The Lorenz number for a specific number of channels $n$ is given by

\begin{multline}
    \mathcal{L}_n = \frac{J_n k_B \beta  }{\delta T G_n}  \\ = \frac{ \left((n-1) \Gamma \left(2+\frac{1}{n}\right)^2-\Gamma \left(1+\frac{1}{n}\right) \Gamma \left(3+\frac{1}{n}\right)\right) \Gamma \left(\frac{3}{2}+\frac{1}{n}\right)}{n \Gamma
   \left(4+\frac{2}{n}\right) \Gamma \left(1+\frac{1}{n}\right)}  \\\times   \frac{ 2^{2+\frac{2}{n}} \pi^{\frac{3}{2}}  k_B^2 }{e^2  } \\
   =  \frac{2+n }{2+3 n} \frac{ \pi^{2}  k_B^2 }{e^2  } = 3 \frac{2+n }{2+3 n} \mathcal{L}_\infty.
\end{multline}

\end{document}

%% file: preamble.tex
\usepackage{geometry}
\usepackage[utf8]{inputenc}

\usepackage{amsmath,amssymb}
\usepackage{bbm}
\usepackage{bm}

\usepackage{graphicx}
\usepackage[dvipsnames]{xcolor} 
\usepackage{subfig}
\usepackage[export]{adjustbox}

\usepackage{hyperref}
\usepackage[capitalise]{cleveref} 

\usepackage{comment}